# Immutable Log Storage as a Service

William Pourmajidi[1], Lei Zhang[1], John Steinbacher[2], Tony Erwin[3], and Andriy Miranskyy[1]

[1]Department of Computer Science, Ryerson University, Toronto, Canada
[2]IBM Canada Lab, Toronto, Canada
[3]IBM Watson and Cloud Platform, Austin, USA

{[1]william.pourmajidi, [2]leizhang, [5]avm}@ryerson.ca, [3]jstein@ca.ibm.com, [4]aerwin@us.ibm.com

*Abstract*—Logs contain critical information about the quality of the rendered services on the Cloud and can be used as digital evidence. Hence, we argue that the critical nature of logs calls for immutability and verification mechanism without a presence of a single trusted party. In this paper, we propose a blockchain-based log system, called Logchain, which can be integrated with existing private and public blockchains. To validate the mechanism, we create Logchain as a Service (LCaaS) by integrating it with Ethereum public blockchain network. We show that the solution is scalable (being able to process 100 log files per second) and fast (being able to "seal" a log file in 23 seconds, on average).

## I. Introduction

In the majority of Cloud offerings, there are two parties involved. A Cloud Service Provider (CSP) owns a pool of computing resources and offers them at a predefined price to a Cloud Service Consumer (CSC) via Internet. The CSP uses continuous monitoring to ensure that the current Quality of Service (QoS) provided to the CSC matches with the one in the signed Service Level Agreement (SLA). While the full control over monitoring systems and generated logs allow CSPs to monitor and maintain Cloud services efficiently, it gives them a controversial power over evidential resources that are important to CSCs. That is, logs are generated and stored on a platform, which is built, managed, and owned by a CSP. Hence, CSPs have full permission on all collected logs. Such situations cause many trust related issues.

To address the trust issue of the current Cloud log storage solutions, our objective is to create an immutable log system called Logchain. We choose blockchain as our data storage model for its immutability and support for the storage of any type of data. We also address the main scalability limitation of the blockchain, namely, the number of computational resources that are needed to verify the integrity of each block. To make the Logchain more accessible, we construct a Logchain as a Service (LCaaS) by implementing an API that can be used to interact with the Logchain.

The idea of the Logchain and its detailed design has been presented at the IEEE CLOUD 2018 [1] and the source code of the prototype can be accessed via [2]. This manuscript expands [1] by showing that LCaaS can be implemented on Ethereum (Section III) and by providing performance evaluation in Section IV.

## II. Design of LCaaS

Current blockchain consensus protocols require every node of the network to process every block of the blockchain, hence a major scalability limitation. We overcome this limitation by segmenting a portion of a blockchain and locking-it-down in a block of a higher level blockchain, i.e., we create a two level hierarchy of blockchains. Validating the integrity of a high-level block confirms the integrity of all the blocks of the lower-level blockchain and leads to a reduction of the number of operations needed to validate the chain.

While common key components of blockchains are necessary to implement a blockchain, our prototype requires additional components. We have expanded the basic genesis block concept and introduced absolute genesis block, relative genesis block, terminal block, super block, and super blockchains. These advancements allow the LCaaS to provide the hierarchical structure (depicted in Fig. 1), which improves the scalability of blockchains. The absolute genesis block is placed as the first block of the first circled blockchain while a relative genesis block is placed at the beginning of every subsequent circled blockchain after the first circled blockchain. As a key element of LCaaS, the terminal blocks are added at the end of a blockchain to "close" it and produce a circled blockchain that is capped. Moreover, super blocks (SB) exhibit the features of regular data blocks except that its *data* element stores all of the field of a terminal block of a circled blockchain. Finally, super blockchain is a blockchain where each of its blocks is a SB.

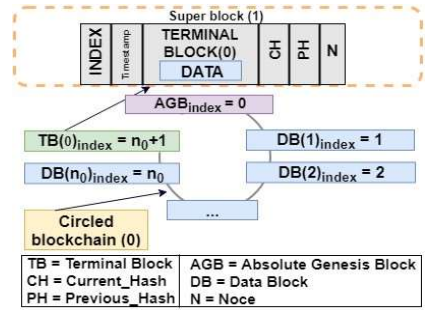

Fig. 1. An SB (with data element) and its relationship with a circled blockchain

TABLE I
THE VALUES OF FACTORS USED IN THE SETUP OF EXPERIMENTS

| Factor | Values |
|---|---|
| Transactions per second ($tps$) | [0.1,1,10,100] |
| Length of circled blockchain ($n$) | [1,10,100] |
| Gas price ($g$) measured in gwei (1 Ether = $10^9$ gwei) | [6,9,20] |
| Number of sample log files | 200 or 1000 |
| Size of sample log files measured in bytes | 64 |

The above novel enhancements allow the LCaaS to provide the hierarchical structure that is needed to overcome scalability limitations of the blockchains.

### III. LCaaS on Ethereum

Given the popularity of Ethereum, we have selected Ethereum as the blockchain platform for integration with the LCaaS. LCaaS is built on top of a private blockchain. In order to replace it with a public blockchain, we propose a composite structure, in which receiving logs and converting them to blocks happens at the LCaaS side and storing the hash encryption and digitally signing them happens over the Ethereum blockchain. Within the Ethereum blockchain, the economics are controlled by an execution fee called gas. The gas is paid by Ether—the Ethereum intrinsic currency [3]. The gas measures the effort (in terms of computational resources) needed to process the transaction. We employ Ethereum test network. Ethereum test networks uses test Ether currency—a virtual Ether that has no monetary value. We use MetaMask Ether Faucet [4] to obtain test Ether. We use Solidity [5] to publish our smart contract. All interactions with the Ethereum blockchain can be traced using Etherscan [6], a web dashboard connected to Ethereum blockchain. Etherscan allows anyone to look up transactions' details by using the sender or recipient address, transaction hash, or block number. An example of a successful transaction of LCaaS on Etherscan can be seen in [7].

### IV. Performance Test and Analysis

To test the performance, we design a load test and run it on our test computer with Intel i7-7500U CPU and 16 GB of RAM. The main goal of the load test is to evaluate the impact of the three configurable factors: incoming transactions (i.e., log files) per second ($tps$), length of circled blockchains ($n$), and gas price ($g$). The values of the factors are given in Tab. I. The permutations of the values of $tps$, $n$, and $g$ (shown in Table I) lead to 36 distinct setups. As for the transaction, we use the digest of a log file (64 bytes long). We use Postman [8] to generate incoming transactions (i.e., log files) to the LCaaS. It is important to mention that one has the option to submit the actual log records or their digest at any desired intervals. We conduct 36 experiments (one for every permutation of the values of the factors listed in Tab. I). To see whether the performance will be affected by $n$, $g$ and $tps$, we perform Pearson and Spearman's correlation analysis, as well as linear regression analysis, on the raw data (i.e., per SB timing) as well as mean, median, and 95th percentile timing of SBs for each experiment. Aggregate statistics are chosen to reduce the amount of noise in the data. We found that none of the factors or the composite factors have any statistically significant relation to the response times, based on the low (< $0.15$) values of correlations and high (> $0.1$) $p$-values of linear models. This implies that the time needed to process SB is dependent mainly on the Ethereum network and availability of the miners. We show a distribution of processing times for SB in Fig. 2. Eyeballing of the distributions suggests that the lower the gas price, the more SB have higher processing time (> 32 seconds), even though the difference is not dramatic. Based on Kolmogorov-Smirnov test, the distribution of $g$ =20 case differs significantly ($p$-value < $0.001$) from the cases when $g$ =9 or $g$ =6. However, the difference between $g$ =6 and $g$ = 9 cases is less pronounced: $p$-value $\approx 0.08$. We were anticipating a stronger difference between all three cases; probably our usage of the test network rather than a production one lead to this anemic difference.

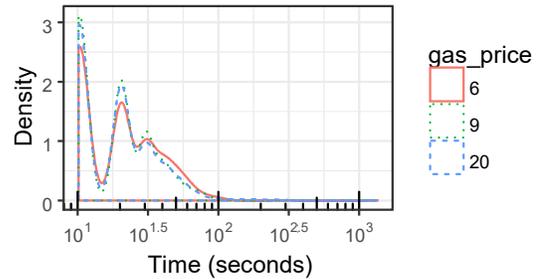

Fig. 2. Density estimate of the processing time for SB

Essentially, our findings show that the network has enough capacity to "absorb" the changes in our workload even in the intense cases, such as $tps$ =100 and $n$ =1. However, in rare cases, the processing time is high: out of 3089 processed SBs, 5 (0.16%) had been processed in between 3 and 5 minutes, and 1 (0.03%) in 23 minutes. As we can see, these cases are rare, but they do exist and we have to be aware of such events.

### V. Conclusions

The proposed solution prevents log tampering, ensuring transparent logging process and establishing trust between all Cloud participants. In the future, we are planning to test LCaaS with other blockchain solutions with the focus on the private blockchain offerings.